\documentclass[superscriptaddress, showpacs,preprintnumbers,amsmath,amssymb]{revtex4}
\usepackage{graphicx}
\usepackage{dcolumn}
\usepackage{bm}
\usepackage{latexsym}
\begin{document}
\title{100 years of Einstein's theory of Brownian motion:\\ from pollen grains to protein trains{\footnote{Based on the inaugural lecture in the Horizon Lecture Series organized by the Physics Society of I.I.T. Kanpur, in the ''World Year of Physics 2005''. }}}
\author{Debashish Chowdhury}
\affiliation{Department of Physics\\
Indian Institute of Technology\\
Kanpur 208016, India.
}
\date{\today}
\begin{abstract} 
Experimental verification of the theoretical predictions made by Albert 
Einstein in his paper, published in 1905, on the molecular mechanisms 
of Brownian motion established the existence of atoms. In the last 100 
years discoveries of many facets of the ubiquitous Brownian motion has 
revolutionized our fundamental understanding of the role of {\it thermal 
fluctuations} in the exotic structures and complex dynamics exhibited by 
soft matter like, for example, colloids, gels, etc. The domain of Brownian 
motion transcends the traditional disciplinary boundaries of physics and 
has become an area of multi-disciplinary research. Brownian motion finds 
applications also in earth and environmental sciences as well as life 
sciences. Nature exploits Brownian motion for running many dynamical 
processes that are crucial for sustaining life. In the first one-third 
of this article I present a brief historical survey of the initial period, 
including works of Brown and Einstein. In the next one-third  I 
introduce the main concepts and the essential theoretical techniques 
used for studying translational as well as rotational Brownian motions 
and the effects of time-independent potentials. In the last one-third 
of this article I 
discuss some contemporary problems on Brownian motion in time-dependent 
potentials, namely, {\it stochastic resonance} and {\it Brownian ratchet}, 
two of the hottest topics in this area of interdisciplinary research. 
\end{abstract}
                                                                                
                                                                                
\maketitle

\section{\label{sec1}Introduction}

The United Nations has declared the year 2005 as the ``World Year of 
Physics'' to commemorate the publication of the three papers of Albert 
Einstein in 1905 on (i) special theory of relativity, (ii) photoelectric 
effect and (iii) Brownian motion \cite{stachel}. These three papers not 
only revolutionized physics but also provided keys to open new frontiers 
in other branches of science and almost all areas of modern technology. 
In one of these three papers \cite{einstein05}, entitled ``On the 
movement of small particles suspended in a stationary liquid demanded 
by the molecular kinetic theory of heat'', Einstein developed a 
quantitative theory of Brownian motion assuming an underlying molecular 
mechanism. Popular science writers have written very little on this 
revolutionary contribution of Einstein; most of the media attention was 
attracted by his theory of relativity although he received the Nobel 
Prize for his theory of photoelectric effect which strengthened the 
foundation of quantum theory laid down somewhat earlier by Max Planck. 
Is Brownian motion, in any sense, less important than its two more 
glamourous cousins, namely relativity and quantum phenomena?

Before answering this question I would like to draw your attention 
to the fact that each of the three revolutionary papers published 
by Einstein in 1905 is concerned with some {\it extreme conditions} 
characterized by a natural constant. The paper on relativity was 
concerned with extremely {\it fast} moving particles whose speed is 
comparable to that of light in vacuuum (usually denoted by the symbol 
$c$). His paper on the photoelectric effect dealt with quantum phenomena 
that dominate physics of extremely {\it small} particles whose action, 
having a dimension $[ML^2T^{-1}]$, is comparable to the Planck's contant 
(usually denoted by the symbol $h$). Similarly, his paper on Brownian 
motion is relevant for the structures of extremely {\it complex} 
systems where the energies associatd with non-covalent bonds are 
comparable to the typical thermal energy $k_BT$, $k_B$ being the 
Boltzmann constant. 

Based on the progress of science and technology over the last 100 
years we can assert that Brownian motion plays important role not 
only in a wide variety of systems studied within the traditional 
disciplinary boundaries of physical sciences but also in systems 
that are subjects of investigation in earth and environmental 
sciences, life sciences as well as in engineering and technology. 
Some examples of these systems and phenomena will be given in this 
article. However the greatest importance of Einstein's theory of 
Brownian motion lies in the fact that experimental verification 
of his theory silenced all skeptics who did not believe in the  
existence of atoms.  

Didn't people believe in the existence of atoms till 1905? Well, 
Greek philosophers like, for example, Democritus and Leucippus 
assumed discrete constituents of matter, John Dalton postulated 
the existence of atoms and, by the end of the nineteenth century 
a molecular kinetic theory of gases was developed by Clausius, 
Maxwell and Boltzmann. Yet, the existence of atoms and molecules 
was not universally accepted. For example, physicist-philosopher 
Ernst Mach believed that atoms have only a didactic utility, i.e., 
they are useful only in deriving experimentally observable results 
while they themselves are purely fictituous. 

The continuing debate of that period regarding the existence of 
atoms has been beautifully summarized in the following words by 
Jacob Brnowski in his {\it Ascent of Man} \cite{bronowski}: 
``Who could think that, only in 1900, peoples were battling, one 
might say to the death, over the issue whether atoms are real or 
not. The great philosopher Ernst Mach in Vienna said, NO. The 
great chemist Wilhelm Ostwald said, NO. And yet one man, at that 
critical turn of the century, stood up for the reality of atoms 
on fundamental grounds of theory. He was Ludwig Boltzmann... The 
ascent of man teetered on a fine intellectual balance at that 
point, because had the anti-atomic doctrines then really won the 
day, our advance would certainly have been set back by decades, 
and perhaps a hundred years.''

Therefore, one must not underestimate the importance of Einstein's 
paper in 1905 on the theory of Brownian motion as it provided a 
testing ground for the validity of the molecular kinetic theory. 
It is an irony of fate that, just when atomic doctrine was on the 
verge of intellectual victory, Ludwig Boltzmann felt defeated and 
committed suicide in 1906. 

This article is organized as follows: in section \ref{sec2} we consider 
the period before Einstein from a historical perspective. In \ref{sec3} 
we study critically Einstein’s original work, followed by the most 
important contributions of some of his contemporaries like Smoluchowski, 
Langevin and others. In his original paper of 1905 Einstein was concerned 
with the translational motion of the center of masses of the Brownian 
particles. Subsequently, rotational Brownian motions of rigid particles as 
well as Brownian shape fluctuations of deformable bodies have been studied 
extensively; some typical examples of these phenomena are given in section 
\ref{sec4}. Mathematical techniques developed for dealing with Brownian 
motion found applications in the noise-driven dynamical phenomena 
involving metastable, bistable and multistable states; these include 
phenomena as diverse as chemical reactions, nucleation of liquid droplets 
in supersaturated vapour, and so on. The general theory of Brownian 
motion in static (i.e., time-independent) external potentials, which 
is applicable to some of these phemomena is briefly discussed in section 
\ref{sec5}. Two of the hottest topics in the area of Brownian motion, 
over the last two decades, are stochastic resonance and Brownian ratchet; 
these two phenomena, which involved Brownian motion in time-dependent 
potentials, are discussed in section \ref{sec6} together with examples 
from not only physical and chemical sciences but also biological sciences 
as well as earth and environmental sciences. This article ends with a brief 
summary and main conclusions given in \ref{sec7}. 

\section{\label{sec2}Period before Einstein}

In 1828 Robert Brown, a famous nineteenth century Botanist, published 
``a brief account of the microscopical observations made in the months of 
June, July and August, 1827 on the particles contained in the pollen 
of plants''. Could the incessant random motion of the particles that he 
observed under his microscope be a consequence of the fact that the 
pollens were collected from living plants? Naturally, he  ``was led next 
to inquire whether this property continued after the death of the plant, 
and for what length of time it was retained.'' He repeated his experiments 
with particles derived not only from dead plants but also from ``rocks of 
all ages,...a fragment of the Sphinx...volcanic ashes, and meteorites from 
various localities''. From these experiments he concluded, ``extremely 
minute particles of solid matter,whether obtained from organic or inorganic 
substances, when suspended in pure water, or in some other aqueous fluids, 
exhibit motions for which I am unable to account...''.

By the time he completed these investigations, he no longer believed the 
random motions to be signatures of life. Following Brown's work, several 
other investigators studied Brownian motion in further detail. All these 
investigations helped in narrowing down the plausible cause(s) of the 
incessant motion of the Brownian particles. For example, temperature 
gradients, capillary actions, convection currents, etc. could be ruled out. 

In the second half of the nineteenth century, Giovanni Cantoni, Joseph 
Delsaulx and Ignace Carbonelle independently speculated that the random 
motion of the Brownian particles was caused by collisions with the molecules 
of the liquid. However, Carl von N\"ageli and William Ramsey argued 
against this possibility. Their arguments were based on the assumption 
that the particle suffered no collision along a linear segment of its 
trajectory except those with two fluid particles at the two ends of the 
segment. If this scenario is true, then, it leads to two puzzles:
(i) how can molecules of water, which are so small compared to the
pollen grain, cause movements of the latter that are large enough
to be visible under an ordinary nineteenth century microscope? \\
(ii) A molecule collides over $10^{12}$ times per second. On the
other hand, our eyes can resolve events that are separated in time
by more than $1/30$ second. Therefore, if each displacement of the
pollen grain is caused by a single collision with a water molecule,
then each such displacement would occur at time intervals of
$10^{-12}$ seconds. But, then, how do our eyes resolve these events
and see them as distinct random displacements of the pollen grain?
On the basis of these arguments N\"ageli and Ramsey tried to rule 
out the mechanism based on molecular collisions. We shall see how 
this paradox was resolved later by Smoluchowski, a contemporary of 
Einstein.

Did Brown really discover the phenomenon which is named after him? No. 
In fact, Brown himself did not claim to have discovered it. On the 
contrary, he wrote ``the facts ascertained respecting the motion of the 
particles of the pollen were never considered by me as wholly original...''. 
Brownian motion had been observed as early as in the fifteenth century by 
Leeuwenhoek, the inventor of optical microscope. Brown critically reviewed 
the works of several of his predecessors and contemporaries on Brownian 
motion. Over the next three quarters of the nineteenth century, many 
investigators studied this phenomenon and speculated on the possible 
underlying mechanisms, major contributors being Gouy and Exner. 
Nevertheless, this phenomenon was named after Brown; this reminds us 
Stigler’s law of eponymy: ``No scientific discovery is named after its 
original discoverer''. 

\section{\label{sec3}Einstein and the theory of Brownian motion}

For the sake of simplicity, we shall write all the equations for 
Brownian motion in one-dimensional space; generalizations to higher 
dimensions is quite straightforward.

\subsection{Einstein} 

Einstein published five papers {\it before 1905} \cite{gearhart}. All of 
these five papers were, in Kuhn's terminlogy, ``normal science''. However, 
the last three of these, which were attempts to address some fundamental 
questions on the molecular-kinetic approach to thermal physics, prepared 
him for the ``scientific revolution'' he created through his paper of 1905 
on Brownian motion \cite{einstein05}. The title of that paper, ``On the movement 
of small particles suspended in a stationary liquid demanded by the 
molecular kinetic theory of heat'', did not even mention Brownian motion !!
Einstein was aware of the possible relevance of his theory in Brownian 
motion but was cautious. He wrote, “it is possible that the movements to be 
discussed here are identical with the so-called ‘Brownian molecular motion’; 
however, the information available to me regarding the latter is so lacking 
in precision,that I can form no judgment in the matter.”

Einstein formulated the problem as follows: ``We must assume that the 
suspended particles perform an irregular movement- even if a very slow 
one- in the liquid, on account of the molecular movement of the liquid''. 
This  is, indeed, a clearly stated assumption regarding the mechanism 
of the irregular movement. 

The main result of Einstein's paper of 1905 on Brownian motion can be 
summarized as follows: the mean-square displacement $<x^2>$ suffered 
by a sphereical Brownian particle, of radius $a$, in time $t$ is given by 
\begin{equation}
<x^2> = \biggl(\frac{RT}{3 \pi N_{av} a \eta}\biggr) t 
\label{eq-meansq}
\end{equation}
where $\eta$ is the viscosity of the fluid, $R$ is the gas constant and 
$N_{av}$ is the Avogadro number. Since $<x^2>$, $t$, $a$ and $\eta$ are 
measurable quantities, the Avogadro number can be determined by using 
the equation (\ref{eq-meansq}).    

Einstein had clear idea of the orders of magnitude that would make the 
movements visible under a microscope. He wrote, ``In this paper it will 
be shown that according to the molecular-kinetic theory of heat, bodies 
of microscopically-visible size suspended in a liquid will perform 
movements of such magnitude that they can be easily observed in a 
microscope, on account of the molecular motions of heat''. Taking an 
explicit example of a spherical Brownian particle of radius one micron, 
he showed that the root-mean-square displacement would be of the order 
of a few microns when observed over a period of one minute. 

Two intermediate steps of his calculation in this paper are also 
extremely important. First, he obtained 
\begin{equation}
\gamma D = k_B T = R T/N_{av}
\label{eq-fdtheorem}
\end{equation}
where $\gamma$ is the coefficient of viscous drag force, $D$ is the 
diffusion constant and $T$ is the temperature. Note that $D$ is a 
measure of the fluctuations in the positions of the Brownian particle 
while $\gamma$ is a measure of energy dissipation; therefore, the 
formula (\ref{eq-fdtheorem}) is a special case of the more general 
theorem, called fluctuation-dissipation  theorem, which was derived 
half a century later. 

The second important result was his derivation of the diffusion equation 
\begin{equation}
\frac{\partial P}{\partial t} = D \frac{\partial^2 P}{dx^2}
\label{eq-diffn}
\end{equation} 
for $P(x,t)$, the probability distribution of the positions of the 
Brownian particle at time $t$. Although diffusion equation was widely 
used already in the nineteenth century in the context of continuum 
theories, Einstein's derivation established a link between the random 
walk of a single particle and the diffusion of many particles.

\begin{figure}[h]
\begin{center}
\includegraphics[angle=-90,width=0.5\columnwidth]{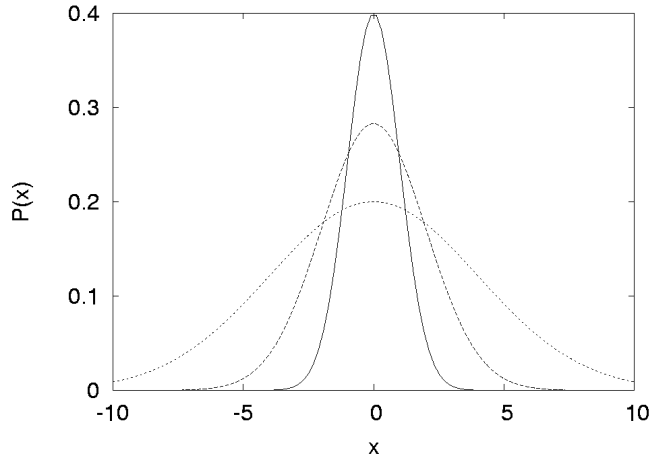}
\end{center}
\caption{The Gaussian probability distribution of a free Brownian particle, 
located initially at $x = 0$, is plotted at three later times; the width  
increases following the equation (\ref{eq-meansq}). 
(Copyright: Indrani Chowdhury; reproduced with permission).
}
\label{fig-gausses}
\end{figure}

For the initial condition $P(x,0) = \delta(x)$, the solution of the 
diffusion equation (\ref{eq-diffn}) is given by 
\begin{equation}
P(x,t) = \frac{1}{[2 \pi \sigma(t)]^{1/2}} e^{-x^2/(2 \sigma^2)} 
\end{equation} 
The root-mean-square displacement $<x^2>$, which corresponds to the width 
of the Gaussians shown in fig.\ref{fig-gausses}, is proportional to 
$\sqrt{t}$.

Einstein's 1905 paper on Brownian motion was not the only paper he 
wrote on this topic. In fact, in the opinion of leading historians of 
science, Einstein's Ph.D.  thesis, which was published in 1906, is 
perhaps more important contribution to the theory of Brownian motion 
than his 1905 paper. But a detailed discussion of his later papers 
on this subject is beyond the scope of this article.

Einstein also realized what would be the fate of kinetic theory in case 
experimental data disagreed with his predictions. he wrote, ``...had 
the prediction of this movement proved to be incorrect, a weighty argument 
would be provided against the molecular-kinetic conception of heat''. 

In 1900 Louis Bachelier's thesis entitled ``Therie de la Speculation''
was examined by three of the greatest mathematician and mathematical
physicists, namely, Paul Appell, Joseph Boussenesq and Henri Poincare.
It was Poincare who wrote the report on that thesis which may be 
regarded as the pioneeing work on the application of mathematical 
theory of financial markets. In his thesis Bachelier postulated that 
stock prices execute Brownian motion and he developed a mathematical 
theory which was, at least in spirit, very similar to the theory 
Einstein developed five years later!

\subsection{Smoluchowski} 

Unlike Einstein, Marian Smoluchowski was familiar with the literature 
on the experimental studies of Brownian motion. If he had not waited 
for testing his own theoretical predictions, the credit for developing 
the first theory of Brownian motion would go to him. He developed the 
theory much before Einstein but he decided to publish it only after he 
saw Einstein's paper which contained similar ideas. In his first paper, 
\cite{smolu06} published in 1906, Smoluchowski also pointed out the 
error in the N\"ageli-Ramsey objection against the original 
Cantoni-Delsaulx-Carbonelle argument. He clarified that each of the 
apparently straight segments of the Brownian trajectory is caused not 
by a single collision with a fluid particle, but by an enormously large 
number of successive kicks it receives from different fluid particles 
which, by rare coincidence, give rise to a net displacement in the same 
direction.  

\subsection{Perrin} 

Jean Perrin, togther with his students and collaborators embarked on 
the experimental testing of Einsteins theoretical predictions. Their 
first task was to prepare a colloidal suspension with dispersed 
particles of appropriate size. They used gamboge, a gum extract, which 
forms spherical particles when dissolved in water. With the samples 
thus prepared, Perrin not only confirmed that the root-mean-square 
displacement of the dispersed particles grow with time $t$ following 
the square-root law (\ref{eq-meansq}) but also made a good estimate of 
the Avogadro number. Einstein himself was surprised by the high level 
of accuracy achieved by Perrin and in a letter to Perrin he admitted 
``I did not believe that it was possible to study the Brownian motion 
with such a precision''. It is true that the critics of molecular 
reality were silenced not by just one set of experiments of Perrin, 
but by the overwhelming evidence that emerged from almost identical 
estimates of the Avogadro number obtained by using many different 
methods. For his outstanding contribution, Jean Perrin was the Nobel 
prize in 1926.

\subsection{Langevin} 

In the {\it first approximation}, we
can approximate the fluid by a {\it continuum}. Therefore, the
classical equations of motion of the Brownian particle can be written
as
\begin{equation}
dx/dt = v
\label{eq-lan1}
\end{equation}
\begin{equation}
M(dv/dt) = F_{ext} - \Gamma v
\label{eq-lan2}
\end{equation}
where, at this level of description, $\Gamma$ is treated as a
phenomenological parameter. One interesting feature of this equation 
is that, in case of tiny particles whose inertia (i.e., mass) is 
negligibly small, $\Gamma v \propto F_{ext}$, a relation reminiscent 
of Aristotelian mechanics which was based on the assumption that 
it is the velocity (rather than the acceleration) which is 
proportional to the external force acting on it.

However, on the scale of the size of a real Brownian particle the
fluid does not appear to be a continuum. In fact, a Brownian particle
``sees'' that the fluid is made of molecules that constantly, but
{\it discretly}, strike this Brownian particle, {\it accelerating}
and {\it decelerating} it perpetually. ``We witness in Brownian
movement the phenomenon of molecular agitation on a reduced scale
by particles very large on molecular scale '' \cite{chandrasekhar}.
A single collision has very small effect on the Brownian particle;
the Brownian motion observed under a microscope is the {\it cumulative
effect of a rapid and random sequence of large number of weak
impulses}. Since the number of collisions suffered by the Brownian
particle is very large, we do not intend to follow its path in any
detail. Instead, we would like to have a {\it stochastic} description
of its movement.
                                                                                
Since equation (\ref{eq-lan2}) is a good first approximation, we
{\it assume} that equation (\ref{eq-lan2}) correctly describes the
{\it average} motion. We now incorporate the effects of the discret
collisions in a stochastic manner by {\it adding} a {\it fluctuating}
force (with vanishing mean) to the frictional force term:
\begin{equation}
M (dv/dt) = F_{ext} - \Gamma v + F_{br}(t)
\label{eq-lan3}
\end{equation}
So far as the ``fluctuating force'' (``noise'') $F_{br}(t)$ is concerned,
we {\it assume}:\\
(i) $F_{br}(t)$ is independent of $v$, and \\
(ii) $F_{br}(t)$ varies {\it extremely rapidly} as compared to the
variation of $v$. Since ``average motion'' is still assumed to be
governed by the equation (\ref{eq-lan2}), we must have
\begin{equation}
<F_{br}(t)> = 0.
\label{eq-noi1}
\end{equation}
Moreover, the assumption (ii) above implies that during small time
intervals $\Delta t$, $v$ and $F_{br}$ change such that $v(t)$ and
$v(t+\Delta t)$ differ infinitesimally but $F_{br}(t)$ and
$F_{br}(t+\Delta t)$ have no correlation:
\begin{equation}
<\xi(t) \xi(t')> = 2 D T \delta(t-t')
\label{eq-noi2}
\end{equation}
where, $\xi = F_{br}/M$ and, at this level of description, $D$ is a 
phenomenological paranmeter. The prefactor $2$ on the right hand side 
of equation (\ref{eq-noi2}) has been chosen for later convenience. 
In fact, soon we shall see that, in order that the Brownian particle 
is in thermal equilibrium with the surrounding fluid, the constant 
$D$ cannot be arbitrary; only a specific choice of $D$ guarantees the 
approach to the appropriate equilibrium Gibbsian distribution.
                                                                                
Note that the spectral density
\begin{equation}
S(\omega) = 2 \int_{-\infty}^{\infty} e^{-i \omega t} <\xi(\tau) \xi(\tau+t)> dt
\end{equation}
implies that, if the noise satisfies the condition (\ref{eq-noi2}),
then $S(\omega) = 4 D$, independent of $\omega$. Since $S(\omega)$
is independent of the frequency $\omega$, this specific form of noise
is called ``white''. In more general cases, the spectral density of
the noise would depend on the frequency $\omega$ and such noises are
called ``colored''. In the simplest formulations of the Langevin
theory of Brownian motion, one {\it assumes} that $\xi(t)$ is
{\it Gaussian distributed} (with vanishing mean) and with correlations
of the form (\ref{eq-noi2}); for obvious reasons, such noises are
referred to as ``Gaussian white noise''. There are some advantages of 
the Gaussian approximation. But these are too technical to be discussed 
here. 
                                       
What is the operational meaning of the symbol $<.>$ of averaging?
The averaging is to be carried out over the distribution of the noise.
This can be implemented practically in two alternative, but equivalent,
ways:\\
{\it either} averaging over an ensemble of many systems consisting of
a single Brownian particle in a surrounding fluid, {\it or} averaging
over a number of Brownian particles in the same fluid, provided they
are sufficiently far apart (possible at low enough density of the
particles) so as not to influence each other.
                                                                                
What is meant by the term ``solution'' of a stochastic equation like
the Langevin equation? Suppose, we observe a Brownian particle under
a microscope over a sufficiently long time interval $0 \leq t \leq T$
and obtain a record of its position $\vec{r}(t)$ as a function of
time $t$. If the observations are made repeatedly, say $N$ times, we
get $N$ trajectories
$$ \vec{r}_1(t), \vec{r}_2(t),...\vec{r}_N(t).$$
In general, these trajectories are all different, i.e., for a given
$t = t^{\ast}$,
$\vec{r}_1(t^{\ast}), \vec{r}_2(t^{\ast}),...\vec{r}_N(t^{\ast})$
are all different from each other. In other words, the motion of the
Brownian particle is not reproducible and, therefore, not deterministic.
Then, what can physics predict about Brownian motion on the basis of
the Langevin equation? Since, we are unable to make deterministic
predictions we make probabilistic ones.
                                                                                
If we repeat the observations a large number of times,we should be
able to find empirically the distribution of $\vec{r}(t)$. In other
words, we can calculate the probability $P(\vec{r},t;\vec{r}_0,\vec{v}_0)$,
which is the probability of finding the particle at position $\vec{r}$
at time $t$, given that its initial position and velocity were
$\vec{r}_0$ and $\vec{v}_0$, respectively. Moreover, we can also
calculate more detailed probability distributions like, for example,
$P(\vec{r},\vec{v},t;\vec{r}_0,\vec{v}_0)$. However, we shall first
look at the moments of these distributions, e.g.,
$<\vec{v}(t)>$, $<\vec{r}^2(t)>$ by using the statistical properties
of noise.
                                                                                
Calculation of the mean-square displacement, with the given initial
position $x = 0$ at $t = 0$, leads to the final result (I leave it as 
an exercise for the students to go through the steps of the calculation)
\begin{equation}
<x^2> = \biggl(\frac{2k_B T}{\Gamma}\biggr) \biggl[t - \biggl(\frac{1}{\gamma}\biggr) \biggl(1-e^{-\gamma t}\biggr)\biggr].
\end{equation}
Let us examine the two limiting cases. When $t \ll \gamma^{-1}$,
\begin{equation}
<x^2> \simeq (k_B T/M) t^2.
\end{equation}
On the other hand, when $t \gg \gamma^{-1}$,
\begin{equation}
<x^2> \simeq (2k_B T/\Gamma) t.
\label{eq-msqlan}
\end{equation}
Thus, the Brownian particle moves, effectively, ``ballistically''
for times $t \ll \gamma^{-1}$ whereas for times $t \gg \gamma^{-1}$
it moves ``diffusively'' with the effective diffusion coefficient
$D = k_B T/\Gamma$. Note that the equation (\ref{eq-msqlan}) is 
identical to the the equation (\ref{eq-meansq}) derived earlier by 
Einstein through his diffusion equation approach.

Thus, the Langevin equation
(\ref{eq-lan3}) is a {\it stochastic} dynamical equation that
accounts for {\it irreversible} processes. On the other hand, in
principle, one can write down the equations of motion for the
Brownian particle as well as that of all the other particles
constituting the heat bath; each of these Hamilton's canonical
equation of motion will not only be {\it deterministic} but will also
exhibit {\it time-reversal symmetry}. Note that, in the Langevin
approach, one writes down only the equation (\ref{eq-lan3}) for
the Brownian particle and does not explicitly describe the dynamics
of the constituents of the heat bath. Therefore, a fundamental question
is: how do the viscous damping term (responsible for irreversibility)
and the random force term (which gives rise to the stochasticity)
appear in the equation of motion of the Brownian particle when one
``projects out'' the degrees of freedom associated with the bath
variables and observes the dynanics in a tiny subspace of the full
phase space of the composite system consisting of the Brownian
particle + Bath?

To my knowledge, the simplest derivation of the stochastic Langevin 
equation for a Brownian particle, starting from the mutually coupled 
deterministic Hamilton's equations (which are equivalent to Newton's 
equation) for the Brownian particle and the molecules of the fluid, 
was given by Robert Zwanzig \cite{zwanzig}. For the simplicity of 
analytical calculations, he modelled the heat bath as a collection of 
harmonic oscillators each of which is coupled to the Brownian particle.
The differential equations satisfied by the position $Q$ and the 
momentum $P$ of the Brownian particle have the general form 
\begin{equation}
\dot{Q} = {P}/{M}
\label{eq-Qdot}
\end{equation}
\begin{equation}
dP/dt = F_{ext} + f(Q(t),\{q(t)\},\{p(t)\})
\label{eq-Pdot}
\end{equation}
where $F_{ext}$ is the external force (not arising from the reservoir) 
while $\{q(t)\}$ and $\{p(t)\}$ denote all the positions and momenta of the 
harmonic oscillators constituting the reservoir. Similarly, one can 
also write down the equations of motion for each of the harmonic 
oscillators constituting the reservoir.

In principle, one can formally integrate the equations of motion for 
the bath variables, in terms their corresponding initial conditions, 
and substitute the formal solutions into the equation (\ref{eq-Pdot}) 
for the Brownian particle. Even at this stage, the resulting equation 
is purely deterministic. But, it involves the intial positions and 
initial momenta of all the harmonic oscillators and, in practice, 
it is impossible to specify such a large number of initial conditions 
exactly. If we now assume that only statistical properties of these 
initial conditions of the bath variables are known (or, postulated) 
we get a differential equation for $P$ which is a slight generalization 
of the Langevin equation (\ref{eq-lan3}). This simple analytical 
calculation demonstrates how both the dissipative viscous drag term 
and the noise term appear in the equation of motion of the Brownian 
particle when the bath degrees of freedom are projected out. 

Thus, the molecules in the fluid medium which give the random ``kicks'' 
to the Brownian particle are also responsible for its  energy 
dissipation because of viscous drag. Therefore, it should not be 
surprising that these two manifestations of the fluid medium are 
related through the Einstein relation (\ref{eq-fdtheorem}). This 
also answers one of puzzles faced by early investigators: in the 
absence of any force imposed on the Brownian particle from outside 
the fluid, why doesn't it come to a complete halt in spite of the 
viscous drag ? The incessant random motion of the Brownian particle 
is maintained for ever by the delicate balance of the random kicks 
it gets from the fluid particles and the energy it dissipates back 
into the fluid via viscous drag.

\subsection{Fokker-Planck versus Langevin approach} 

Einstein's approach has been generalized by several of his contemporaries 
including Fokker, Planck, Smoluchowski and others. This general theoretical 
framework 
is now called the Fokker-Planck approach \cite{fpleqn}. In this approach, 
one deals with a {\it deterministic} partial differential equation for a 
probability density. For example, suppose 
$P(\vec{r},\vec{v};t|\vec{r}_0,\vec{v}_0)$ 
be the conditional probability that, at time $t$, the Brownian particle is 
located at $\vec{r}$ and has velocity $\vec{v}$, given that its initial 
(i.e., at time $t = 0$) position and velocity were $\vec{r}_0,\vec{v}_0$. 
Since the total probability integrated over all space and all velocities 
is conserved (i.e,, does not change with time), the probability density 
$P$ satisfies an equation of continuity 
\begin{equation}
\frac{\partial P}{\partial t} + \frac{\partial J_p}{\partial x} = 0
\end{equation}
where $J_p$ is the corresponding probability current. Note that the 
probability density and the probability current density are analogs 
of the electrical charge density and electrical current density in 
electrodynamics where it follows from the conservation of electrical 
charge in the system.

The probability current density $J_p$ gets contributions from two 
sources: the diffusion current, given by Fick's law, is caused by 
the concentration gradient, while the drift current is imposed by 
the externally applied force $F$. Thus, 
\begin{equation}
J_p(x,t) = - D \biggl(\frac{\partial P}{\partial x}\biggr) + \biggl(\frac{F}{\Gamma}\biggr) P 
\label{eq-probcurr}
\end{equation}
where the first and the second terms on the right hand side arise from 
diffusion and drift, respectively. The expression (\ref{eq-probcurr}) 
can be recast in several alternative, but equivalent, forms using 
the relation between the force $F$ and the corresponding potential 
$U$, namely, $F = - dU/dx$ and the Einstein relation $\Gamma D = k_B T$.

In contrast, the Langevin approach \cite{langevin} is based on a 
{\it stochastic} differential equation for the individual Brownian 
particle and is, in spirit, closer to Newton's equation. Because of 
the stochasticity, unique initial condition does not lead to a unique 
trajectory of the particle. 

Stochasticity can enter into a differential equation either as an 
additive term or as a multiplicative factor.  For example, the 
Langevin equation for a Brownian Harmonic oscillator in one-dimension 
is given by 
\begin{equation}
M \frac{d^2X}{dX^2} = - M \omega^2 X - \Gamma \frac{dX}{dt} + F_B(t) 
\end{equation}
where the random Brownian force $F_B(t)$ introduces stochasticity 
that appears as an {\it additive term} in the dynamical equation. 
In contrast, the Langevin equation for the so-called Kubo oscillator 
is given by 
\begin{equation}
M \frac{d^2X}{dX^2} = - M \omega_B^2(t) X - \Gamma \frac{dX}{dt}  
\end{equation}
where frequency $\omega_B(t)$ is random and, thus, stochasticity 
enters into the dynamical equation as a {\it multiplicative factor}.


\section{\label{sec4}Beyond translation- rotation and shape fluctuations}

In undergraduate mechanics courses in colleges (or universities), normally, 
a student first learns Newtonian mechanics of point particles which 
also describes the motion of the center of mass of extended objects. 
Then, one learns the deal with the rotational motion of rigid bodies. 
Finally, a student is exposed to the mechanics of deformable bodies, 
i.e., elastic solids and fluids. 
So far in this article we have considered only the translational motion 
of the center of mass of the Brownian particles. In this section we 
shall consider rotational Brownian motion of rigid particles and the 
shape fluctuations of soft materials caused by the Brownian motion of 
these deformable bodies.

\subsection{Rotational Brownian motion of rigid bodies}

To my knowledge, one of the earliest direct experimental observations of 
the rotational Brownian motion was made by Gerlach \cite{gerlach} 
using a tiny mirror fixed on a very fine wire; some of the fundamental 
questions on this problem were addressed theoretically soon thereafter 
by Uhlenbeck and Goudsmit \cite{uhlenrot}. 
This is a relatively simple problem because the rotation 
involves only a single angle $\theta$ which measures the angular deflection. 
The corresponding Langevin equation has the form 
\begin{equation}
I \biggl(\frac{d^2\theta}{dt^2}\biggr) = {\cal T}_{ext} - G \theta - \alpha \biggl(\frac{d\theta}{dt}\biggr) + {\cal T}_{br}
\end{equation}
where $I$ is the moment of inertia of the oscillator, $\alpha$ is the 
friction coefficient, $G$ is the torsional elastic constant of the 
fiber, ${\cal T}_{ext}$ is the external torque and ${\cal T}_{br}$ 
is the Brownian (i.e., random) torque. Each term of this Langevin 
equation is the rotational counterpart of the corresponding term in 
the Langevin equation (\ref{eq-lan3}) for translational Brownian motion. 

Interestingly, three quarters   
of a century later the problem of rotational Brownian motion of a mirror 
was reinvestigated by replacing air by a fluidized granular medium. 
In this novel experiment \cite{granular} the torsion oscillator was 
immersed in a container filled with glass beads and the noisy vertical 
vibration of the container took place at frequencies much higher than 
the natural frequency of the torsion oscillator.

The Langevin equation for the more general cases of rotation of a 
rigid body in three dimensions has more complex form. Recall that 
the rotational motion of a macroscopic asymmetrical object is given 
by the Euler equation. The corresponding Euler-Langevin equation for 
rotational Brownian motion has the general form 
\begin{equation}
\frac{d\vec{L}}{dt} + \vec{\omega} \times \vec{L} = {\cal T}_{ext} - \Gamma \omega + {\cal T}_{br}(t) 
\label{eq-eulangevin}
\end{equation}
with $\vec{L} = I \omega$, 
where $I$ is the moment of inertia of the body and $\omega$ is its 
angular velocity; ${\cal T}_{ext}$ is the externally imposed torque 
while ${\cal T}_{br}(t)$ is the random noise torque. For the sake of 
simplicity one often assumes a Gaussian white noise torque ${\cal T}_{br}(t)$.

How should we write down the Fokker-Planck equation corresponding to 
the Euler-Langevin equation (\ref{eq-eulangevin})? Let us denote by 
the symbol $S'$ a principal coordinate system fixed in the rigid body. 
The orientation of the body can be specified by the Euler angles 
$\Omega_1,\Omega_2,\Omega_3$ of $S'$ with respect to the laboratory 
coordinate system $S$. We can now define the probability density 
$P(\Omega_1(t),\Omega_2(t),\Omega_3(t);\omega_1(t),\omega_2(t),\omega_3(t)|\Omega_1(0),\Omega_2(0),\Omega_3(0);\omega_1(0),\omega_2(0),\omega_3(0))$ which 
represents the conditional probability that, at time $t$, the Euler angles  
are $\Omega_1(t),\Omega_2(t),\Omega_3(t)$ and the angular velocities $\omega_1(t),\omega_2(t),\omega_3(t)$, given that 
the corresponding initial values were $\Omega_1(0),\Omega_2(0),\Omega_3(0)$ 
and $\omega_1(0),\omega_2(0),\omega_3(0)$, respectively.

The problem of rotational Brownian motion of a sphere was briefly 
mentioned in a paper published by Einstein in 1906 \cite{einstein06}. 
Investigaton of the rotational Brownian motion in the context of 
dielectric relaxation was initiated by Peter Debye and extended 
by many authors in the second half of the twentieth century \cite{conel}.

Another related problem is the relaxational dynamics of large 
single-domain particles in rocks \cite{rock}. Each of these particles 
consists of a large number of individual moments all aligned parallel 
to each other such that the particle posseses a giant magnetic moment. 
Since the particle is embedded in a solid matrix, it cannot rotate 
physically but the direction of the magmetic moment can undergo 
Brownian rotation. A collection of such single-domain particles will 
be aligned parrel to the externally applied magnetic field. Then, 
after the field is switched off, the remanent magnetization $M_r$ 
will vanish as  
\begin{equation}
M_r = M_s e^{-t/\tau},  
\end{equation}
where $M_s$ is the magnetization of a non-relaxing particle, $t$ 
is the time elapsed after the field is switched off and  
\begin{equation}
\tau = \tau_0 e^{A V/k_BT} 
\end{equation}
is the relaxation time where $V$ is the volume of the particle and 
$\tau_0 \sim 10^{-9}$ sec. Therefore, varying $V$ and/or $T$, the 
relaxation time $\tau$ can be made to vary from $10^{-9}$ sec. to 
millions of years. 

It was pointed out by Louis Neel that, at a given temperature $T$, 
the particle magnetization will appear ``blocked'' (i.e., frozen 
in time) in any dynamic experiment where the frequency of the 
measurement $\omega_m$ is such that $\tau \gg \omega_m^{-1}$. 

Blocking of the magnteization of the super-paramagnetic particles 
finds important applications in paleomagnetism (geomagnetism), 
as the history of the earth's magnetic field remains frozen in the 
rocks. During the early stage of the formation of the rock at a 
relatively higher temperature, the magnetic particles exist in 
thermal equilibrium with the Earth's magnetic field, but later, 
as the rock cools, the magnetization of these particles get 
``blocked'' and they retain the memory the direction of the Earth's 
magnetic field. 

\subsection{Brownian motion of deformable bodies: shape fluctuations}

A linear polymer is a simple example of a deformable body which is, 
effectively, one dimensional. The Brownian forces acting on such an 
object in aqueous medium can give rise to random {\it wiggling}, i.e., 
random fluctuations in its shape. The random Brownian forces tend to 
induce wiggles in the polymer chain while the bending stiffness tends 
to restore its linear shape. These two competing effects determines 
overall conformation of the polymer chain. One of the most important 
effects of its Brownian wiggling is that, even in the absence of any 
energy cost for creating such wiggles, the polymer behaves, effectively, 
as a spring where its spring constant is temperature-dependent and 
the corresponding restoring force it exerts is of purely entropic origin.

Suppose $\hat{n}(0)$ and $\hat{n}(s)$ are the unit normals to 
the polymer at two points on the polymer separated by a distance 
$s$ measured along the contour of the chain. Then, the correlation 
between the orientations of these two unit normals decreases 
exponentially with increasing $s$, i.e., proportional to 
$exp(- s/\xi_p)$ where $\xi_p$, the {\it persistence length}, is 
determined by the ratio of the bending stiffness energy and the 
thermal energy $k_B T$. If the total length of the polymer is $L$, 
it appears stiff when $L << \xi_p$ whereas it appears floppy when 
$L >> \xi_p$. Microtubules have very long persistence length.

Similarly, Brownian motion of a soft membrane, e.g., the plasma membrane 
of a red-blood cell, manifests as ``flickering'' of the, effectively, 
two-dimensional elastic sheet. The Brownian shape fluctuations of 
soft membranes have many important consequences. For example, 
consider a stack of such membranes which have a tendency to stick to 
each other because of the ubiquitous Van der Waals attractions. 
However, at all non-zero temperatures the Brownian shape fluctuations 
cause the membranes to bump against each other; the higher is the 
temperature the stronger is the, effectively, repulsive entropic 
force. As a consequence of this competition between the two forces, 
an unbinding phase transition takes place in the system at a 
characteristic temperature as the temperature is raised from below.


\section{\label{sec5}Brownian motion in external static potential}

Translational Brownian motion of a particle under the influence of an 
external {\it linear} potential of the form $U(x) = ax$ is relevant, for 
example, in the context of sedimentation of colloidal particles under 
gravity \cite{chandrasekhar}. Brownian motion of a harmonically bound 
particle \cite{uhlenorn}, i.e., a particle subjected to a {\it quadratic} 
potential of the form $U(x) = a x^2$, is a reasonably good model for the 
dynamics of tiny spherical dielectric particle trapped by an optical 
tweezer. In order to satisfy the law of equipartition of energy in 
thermodynamic equilibrium, $<x^2>$ approaches the value $k_BT/(m\omega^2)$ 
in the limit of extremely long time limit; Uhlenbeck and Ornstein 
\cite{uhlenorn} derived the exact expression valid for all times and, 
hence, showed how $<x^2>$ approaches the asymptotic value with the 
passage of time.  

The potential $U(x) = - a x^2 + b x^4$  has two equally deep minima 
which are separated from each other by an energy barrier; Brownian 
motion of a particle subjected to such a potential leads to noise 
assisted transitions, back and forth, from one well to the other. 
The average waiting time $T_K$ between two successive noise-induced 
transitions increases exponentially with the increase of the barrier 
height. Noise-induced transitions in bistable systems have found 
applications in a wide variety of systems; we shall call $T_K$ as 
the Kramers time in honor of Hendrik Kramers who considered such 
problems first in the context of chemical reaction rate theory 
in his classic paper entitled ``Brownian motion in a field of force 
and the diffusion model of chemical reactions'' \cite{hanggi,melnikov}.  

Kramers was not the first to consider noise-induced transitions from 
a potential well. In fact, in 1935, Becker and D\"oring studied 
the problem noise-assisted hopping of a barrier to escape from a 
metastable state. The problem of noise-induced transitions in systems 
with metastable, bistable or multistable systems has a long history 
with abundant examples of unintentional rediscoveries and rederivation 
of results by experts from different disciplines, often using different 
terminologies \cite{landauer}. Nevertheless, this shows the breadth 
of coverage of this multidisciplinary umbrella and the wide range of 
applicability of the concepts and techniques.

\section{\label{sec6}Brownian motion in time-dependent potential}

In the preceeding section we have considered Brownian motion in static 
(time-independent) external potential. However, two of the hottest 
topics in the area of Brownian motion which have kept many physicists 
busy for 
the last quarter of a century, are related to Brownian motion in 
time-dependent potentials. In the following two subsections we briefly 
discuss these two phenomena, namely, stochastic resonance and Brownian 
ratchet.

\subsection{Stochastic resonance and applications}

\begin{figure}[h]
\begin{center}
\includegraphics[width=0.5\columnwidth]{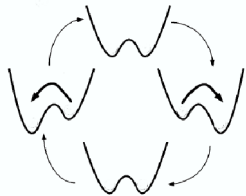}
\end{center}
\caption{The back and forth tilting of the bistable potential in one 
cycle of the periodic focing. 
(Copyright: Indrani Chowdhury; reproduced with permission).
}
\label{fig-stres}
\end{figure}

Let us begin with a Brownian particle subjected to a bistable potential.
Suppose a small amplitude periodic forcing is added so that the left
and the right wells periodically exchange their relative stability as
shown in fig.\ref{fig-stres}. Let $T_{p} = 2 \pi/\Omega_p$ be the time 
period of the periodic forcing. Then, the potential $U(x)$ is given by 
\begin{equation}
U(x,t) = \frac{a}{2} x^2 - \frac{b}{4} x^4 - A_0 x ~cos(\Omega_p t)
\end{equation} 
and the corresponding Langevin equation for the Brownian particle is 
\begin{equation}
m \frac{d^2x}{dt^2} = - a x + b x^3 + A_0 ~cos(\Omega_p t) + F_{br}
\end{equation} 
Note that the periodic forcing is too weak to induce
transition in the position of the particle from one well to the other
without assistance from noise. However, in the presence of noise, even
in the absence of forcing, noise-induced transition from one well to
the other goes on. Now, extending the concept of resonance, we introduce
the concept of stochastic resonence by the condition
\begin{equation}
2 T_K(D) = T_{p}
\end{equation}
where $T_K$ is the Kramers time and it depends on the strength $D$ of
the noise \cite{gammaitoni,moss}.
                                                                                
The phenomenon of stochastic resonance has been demonstrated directly
in laboratory experiments \cite{libchab}. A micron-size dielectric 
bead is used as
the Brownian particle and a bistable potential is created using two
optical (laser) traps. The most important quantity characterizing a  
stochastic resonance is the signal-to-noise (SNR) ratio. The signature 
of a stochastic resonance is that the SNR, which vanishes in the absence 
of noise, rises with the increase of noise intensity and exhibits a 
maximum at an optimum level of noise intensity; on further increase 
of noise intensity SNR decreases because of the randomization caused 
by the noise. In other words, contrasy to naive expectations, noise 
can have a constructive effect in enhancing the signal over an 
appropriately chosen window of noise intensity.
Not surprisingly, it finds applications in electrical engineering. 
Moreover, many organisms seem to use stochatic resonance for sensory 
perception; these include, for example, electro-receptors of paddlefish 
mechano-receptors of crayfish, etc.

Stochastic resosnance has been evoked to explain the periodic occurrence
of Ice age on earth; the period is estimated to be approximately
$100,000$ years. Suppose the ice-covered and water-covered earth
correspond to the two local minima. Eccentricity of the earth's
orbit (and, therefore, incoming solar radiation) varies periodically
with a period of about $T_{p} \simeq 100,000$ years. But, this
variation is too weak to cause the transition from ice-covered to
water-covered earth and vice versa. It has been suggested 
that random noise in the climatic conditions can give rise to a
stochastic resonance causing a transition between the two local
minima with a period of about 100,000 years.

\subsection{Brownian ratchet and applications} 

\begin{figure}[h]
\begin{center}
\includegraphics[width=0.5\columnwidth]{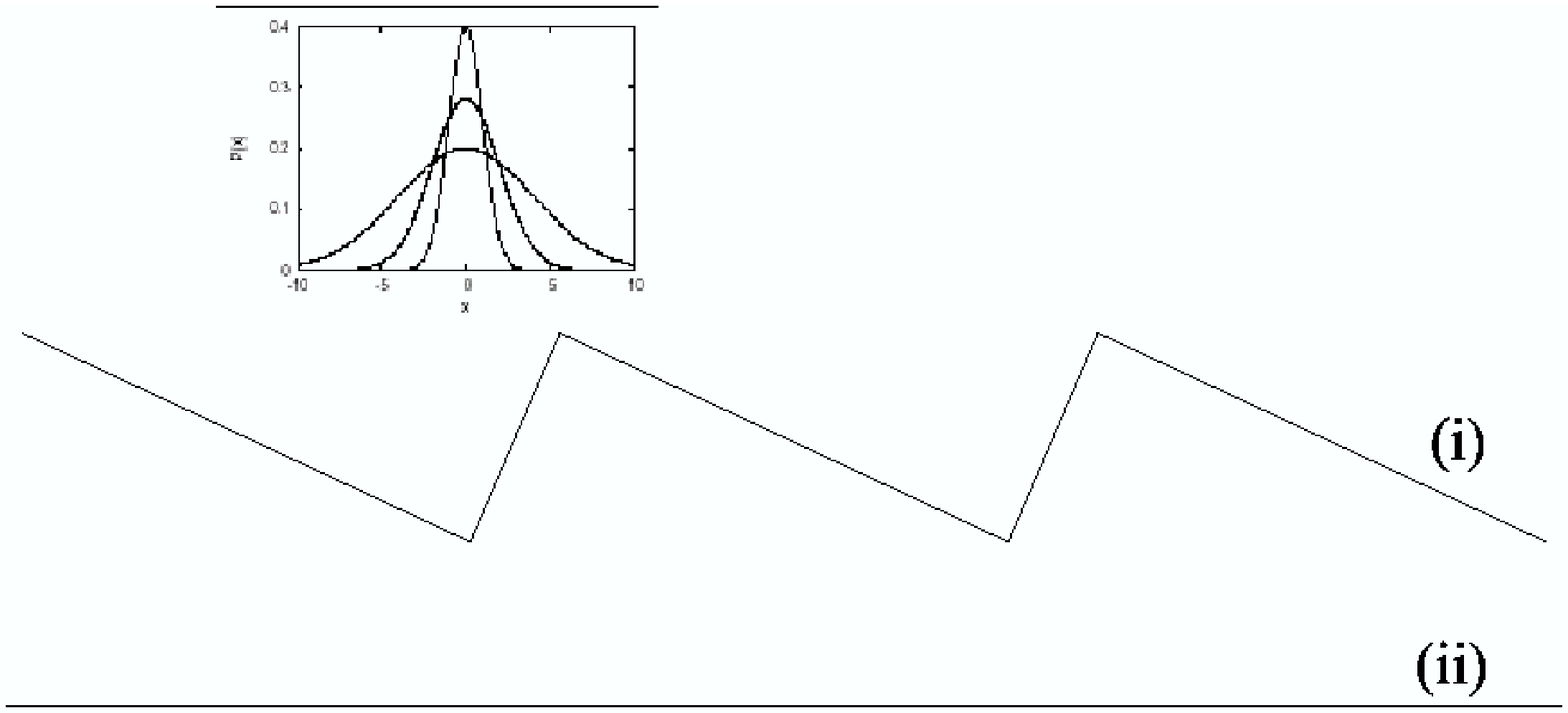}
\end{center}
\caption{The two forms of the time-dependent potential used for 
implementing the Brownian ratchet mechanism.
(Copyright: Indrani Chowdhury; reproduced with permission).
}
\label{fig-ratgauss}
\end{figure}

Let us now consider a Brownian particle subjected to a {\it time-dependent} 
potential, in addition to the viscous drag (or, frictional force). The 
potential switches between the two forms (i) and (ii) shown in 
fig.\ref{fig-ratgauss}. The sawtooth form (i) is spatially {\it periodic} 
where each period has an {\it asymmetric} shape. In contrast, 
the form (ii) is flat so that the particle does not experience any 
external force imposed on it when the potential has the form (ii). 
Note that, in the left part of each well in (i) the particle 
experiences a rightward force whereas in the right part of the same 
well it is subjected to a leftward force. Moreover, the spatially 
averaged force experienced by the particle in each well of length 
$\ell$ is 
\begin{eqnarray}
<F> = - \frac{1}{\ell}\int_0^{\ell} \biggl(\frac{\partial U}{\partial x}\biggr) dx\nonumber \\
= U(0) - U(\ell) = 0 
\end{eqnarray} 
because of the spatially periodic form of the potential (i). What 
makes this problem so interesting is that, in spite of vanishing average  
force acting on it, the particle can still exhibit directed, albeit 
noisy, rightward motion.

In order to understand the underlying physical principles, let us 
assume that initially the potential has the shape (i) and the 
particle is located at a point on the line that corresponds to the 
bottom of a well. Now the potential is switched off so that it makes 
a transition to the form (ii). Immediately, the free particle begins 
to execute a Brownian motion and the corresponding Gaussian profile 
of the probability distribution begins to spread with the passage of 
time.  If the potential is again switched on before the Gaussian profile 
gets enough time for spreading beyond the original well, the particle 
will return to its original initial position. But, if the period 
during which the potential remains off is sufficiently long, so that 
the Gaussian probability distribution has a non-vanishing tail 
overlapping with the neighbouring well on the right side of the 
original well, then there is a small non-vanishing probability that 
the particle will move forward towards right by one period when the 
potential is switched on. 

\begin{figure}[h]
\begin{center}
\includegraphics[width=0.5\columnwidth]{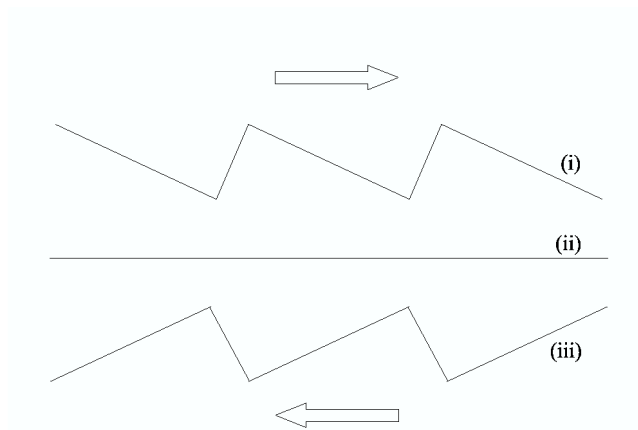}
\end{center}
\caption{The direction of the motion of the particle in a Brownian 
ratchet is determined by the form of the asymmetry of the potential 
in each period. 
(Copyright: Indrani Chowdhury; reproduced with permission).
}
\label{fig-ratbidir}
\end{figure}

In this mechanism, the particle moves forward not because of any  
force imposed on it but because of its Brownian motion. The system 
is, however, not in equilibrium because energy is pumped into it 
during every period in switching the potential between the two 
forms. In other words, the system works as a rectifier where the 
Brownian motion, in principle, could have given rise to both 
forward and backward movements of the particle in the multiples of 
$\ell$, but the backward motion of the particle is suppressed by 
a combination of (a) the time dependence and (b) spatial asymmetry 
(in form (i)) of the potential. In fact, the direction of motion 
of the particle can be revsered by replacing the potential (i) by 
the potential (iii) shown in fig.\ref{fig-ratbidir}.

\begin{figure}[h]
\begin{center}
\includegraphics[width=0.5\columnwidth]{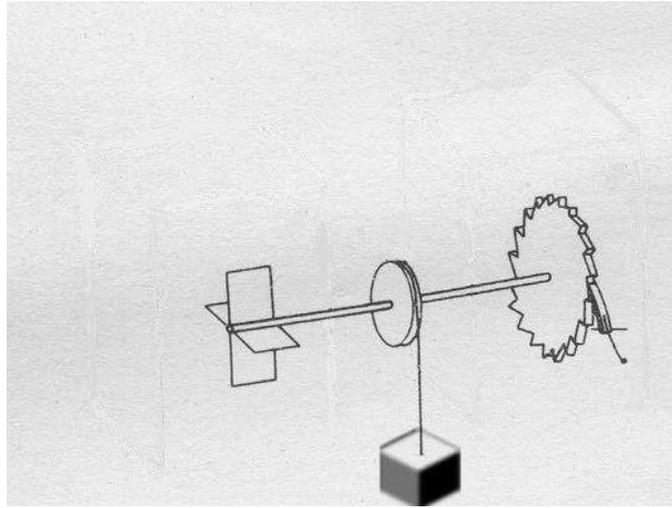}
\end{center}
\caption{Feynman's ratchet and pawl arrangement. 
(Copyright: Indrani Chowdhury; reproduced with permission).
}
\label{fig-ratrot}
\end{figure}

\begin{figure}[h]
\begin{center}
\includegraphics[width=0.5\columnwidth]{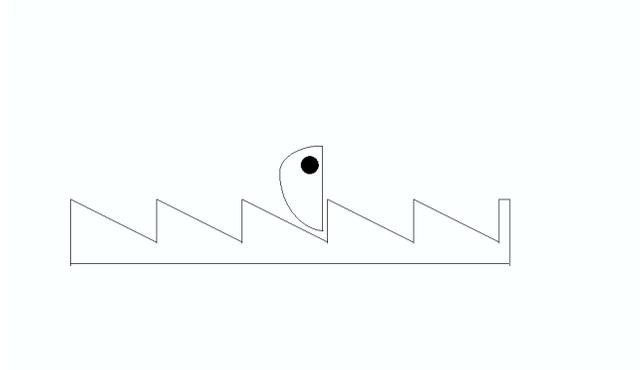}
\end{center}
\caption{A linear ratchet and pawl arrangement. 
(Copyright: Indrani Chowdhury; reproduced with permission).
}
\label{fig-ratlin}
\end{figure}

The mechanism of directional movement discussed above is called a 
Brownian ratchet \cite{reiman} for reasons which we shall now clarify.
The concept of Brownian ratchet was popularized by Feynman through his 
lectures \cite{feynman} although, historically, it was introduced by 
Smoluchowski \cite{smolurat}. Consider the ratchet and pawl arrangement 
shown in fig.\ref{fig-ratrot}. The random bombardment of the vanes by 
the air molecules gives rise to torques which fluctuates randomly both 
in magnitude and direction. Because of the asymmetric shape of each of 
the teeth, it may appear, the ratchet would move countercloclowise more 
easily than clockwise (when viewed from the left side) leading to its directed counterclockwise, albeit 
noisy, rotation. In principle, it should then be possible to exploit 
such directed rotation to perform mechanical work. However, any such 
device, if it really existed, would violate the second law of thermodynamics 
because it would extract thermal energy from its environment, by cooling 
the environment spontaneously, and convert that energy into mechanical 
work. Feynman resolved the apparent paradox by pointing out that both 
the clockwise and counterclockwise rotations are actually equally likely 
because the pawl also executes random Brownian motion because of the 
random extension and compression of the spring that keeps it pressed 
against the wheel of the ratchet.  
A linear design of the Brownian ratchet is shown in fig.\ref{fig-ratlin}

Brownian ratchet has its counterpart in the abstract theory of games. 
In particular, Juan Parrondo \cite{parrondo} proposed a game with two separate rules, 
say $A$ and $B$, of the game. Even in situations where both the rules 
will ruin the gambler, Parrondo showed that the gambler can win by 
using the rules $A$ and $B$ alternately. It is not difficult to map 
this problem onto the Brownian ratchet mechanism depicted in fig.
(\ref{fig-ratgauss}) and the winning of the gambler corresponds to 
the directed movement of the Brownian particle in fig.(\ref{fig-ratgauss}).

The ratcheting via time-dependent potential discussed above it not 
merely a theoretical possibility but nature exploits this for 
driving a class of molecular motors inside cells of living organisms; 
this includes KIF1A, a family of kinesin motor proteins \cite{chowreso}. 
Such molecular motors move along microtubule filaments just as trains 
move along their tracks.
 
A Brownian-ratchet based mechanism has been proposed \cite{simon} for 
translocation of proteins across membranes. This is easy to 
understand using a picture similar to the ratchet shown in the 
fig.\ref{fig-ratlin}. Proteins, are known to 
unfold before translocation through a narrow pore in the membrane. 
Once the tip of the protein successfully penetrates the membrane, 
it can translocate through Brownian motion 
provided there exist some mechanism to rectify its backward movements. 
Several possible mechanisms for such rectification have been proposed 
including binding of chaperonins at designated binding sites along 
the translocated part of the macromolecule \cite{chowreso}. 

ATP is the energy currency of almost all eukaryotic cells and the cell 
synthesizes ATP from the raw materials using a machine, called ATP 
synthase, which is bound to the mitochondrial (chloroplast) membrane of 
animal (plant) cells. To my knowledge, this is smallest among all the 
natural and man made rotary motors. This complex motor actually consists 
of two reversible parts, namely $F_0$ and $F_1$, which are coupled to 
each other. A Brownian-ratchet mechanism has been suggested \cite{oster} 
for the rotary motor $F_0$. Detailed structure and function of this 
natural nano-motor will be considered in a separate article \cite{chowreso}.

\section{\label{sec7}Summary and conclusion}

What started as a curiosity of microscopists, who were baffled by the 
random movements of the pollen grains in water, turned out to be one 
of the most challenging scientific problems that could not be solved 
by anybody till the beginning of the twentieth century. It was Albert 
Einstein who, in one of his three revolutionary papers of 1905, 
published the correct theory of Brownian motion. His theoretical 
predictions were confirmed by a series of experiments on colloidal 
dispersions by Jean Perrin and his collaborators. These investigation 
of Brownian motion in collidal dispersions not only helped in silencing 
the critics of the molecular kinetic theory of matter but also laid 
down the foundation of statistical mechanics.

By the end of the first quarter of the twentieth century quantum theory 
became the darling of the majority of the physicists and the colloidal 
suspensions lost its appeal. Over the next quarter of a century progress 
was rather slow but steady. However, in the second half of the twentieth 
century, motivated partly by the industrial demand for novel materials, 
physicists and engineers discovered great potential of the soft materials 
\cite{frey}, including colloids which gradually regained its past glory 
\cite{haw}. Moreover, revolution in optical microscopy in the last ten 
years has provided a glimpse of the cellular interior, a wonderland 
dominated by Brownian motion. Preliminary explorations in this new 
frontier of research indicate that, instead of being a nuisance, the 
Brownian motion is, perhaps, fully exploited by Nature to its advantage 
not merely to survive but to thrive. Brownian motion of pollen grains does 
not arise from any process of life but some of the least understood 
processes of life, including the train-like motion of the biomolecular 
motors on the filamentary tracks, may not be possible without Brownian 
motion!

\noindent{\bf Acknowledgements:} I thank Manoj Harbola and Ambarish Kunwar 
for a critical reading of the manuscript. 



\begin{thebibliography}{99}
\bibitem{chowreso} Please see the elementary review article by D. Chowdhury, (to be be published).
\end{thebibliography}
\end{document}